\documentclass[prl,groupedaddress,twocolumn,showpacs]{revtex4}
\usepackage[dvips]{graphicx}
\newcommand{\ped}[1]{\ensuremath{_{\rm #1}}}

\begin{document}

\title{Josephson effect in MgB$_2$ break junctions}
\author{R.\@S. Gonnelli}
\email[Corresponding author. E-mail:]{gonnelli@polito.it}
\affiliation{INFM - Dipartimento di Fisica, Politecnico di Torino,
10129 Torino, Italy}
\author{A. Calzolari}
\affiliation{INFM - Dipartimento di Fisica, Politecnico di
Torino, 10129 Torino, Italy}
\author{D. Daghero}
\affiliation{INFM - Dipartimento di Fisica, Politecnico di
Torino, 10129 Torino, Italy}
\author{G.\@A. Ummarino}
\affiliation{INFM - Dipartimento di Fisica, Politecnico di
Torino, 10129 Torino, Italy}
\author{V.\@A. Stepanov}
\affiliation{P.N. Lebedev Physical Institute, Russian Academy of
Sciences, 117924 Moscow, Russia}
\author{G. Giunchi}
\affiliation{Edison S.p.A., Foro Buonaparte 31, 20121 Milano,
Italy}

\author{S. Ceresara}
\affiliation{Edison S.p.A., Foro Buonaparte 31, 20121 Milano,
Italy}

\author{G. Ripamonti}
\affiliation{Edison S.p.A., Foro Buonaparte 31, 20121 Milano,
Italy}

\date{today}

\begin{abstract}
We present the first observation of the DC and AC Josephson effect
in MgB$_{2}$ break junctions. The junctions, obtained at 4.2 K in
high-quality, high-density polycrystalline metallic MgB$_{2}$
samples, show a non-hysteretic DC Josephson effect. By irradiating
the junctions with microwaves we observe clear Shapiro steps
spaced by the ideal $\Delta V$ value. The temperature dependence
of the DC Josephson current and the dependence of the height of
the steps on the microwave power are obtained. These results prove
the conventional nature of the pairs in MgB$_{2}$ and give
evidence of the superconductor-normal metal-superconductor weak
link character of these junctions.

\end{abstract}
 \maketitle

The discovery of superconductivity at 39 K in magnesium diboride
\cite{ref1} has raised great excitement in the scientific
community. Apart from the possibility of finding new simple
compounds with a higher critical temperature, a great interest
arises from the possible applications of this new superconductor.
Among the others we can mention the realization of superconducting
cavities for particle accelerators working at the liquid helium
temperature and superconducting electronics with good performances
at a temperature ($\sim$ 10-15 K) accessible to cryocoolers. These
goals require the realization of high-quality thin films and
junctions of MgB$_2$ but also the knowledge of the fundamental
superconducting properties of the material. Josephson junctions
are the heart of most of the superconducting electronic devices
and, therefore, the knowledge of their DC and AC properties and of
their temperature dependence is crucial. On the other hand, the
Josephson effect is a \emph{direct probe} for the existence of
pairs and a precise instrument for the study of the surface
properties of superconductors. From the $I-V$ characteristics of
Josephson contacts the superconducting gap $\Delta$ can also be
estimated. Many papers have recently appeared in literature
concerning the determination of $\Delta$ from tunneling
\cite{ref2,ref3,ref4,ref5}, Andreev reflection \cite{ref6,ref7},
angle resolved photoemission specroscopy (ARPES) \cite{ref8,ref9},
specific heat \cite{ref10} and Raman \cite{ref11} measurements.
The results are still controversial: $\Delta$ ranges between 2 and
7 meV in the different experiments but its temperature dependence
seems to approximately follow the BCS behaviour \cite{ref6,ref7}.
Various authors have reported the possible presence of a depressed
superconducting layer on the MgB$_2$ surface \cite{ref2,ref7}.

In the present letter we report the first (to our knowledge)
observation of the DC and AC Josephson effect in MgB$_2$ junctions
obtained by using the break-junction technique. The measured $I-V$
curves show values of the product $I_\mathrm{c}R_\mathrm{N}$
(where $I_\mathrm{c}$ is the critical current and $R_\mathrm{N}$
is the normal resistance) which are smaller than the value
predicted by the BCS theory, and a reduced $T_\mathrm{c}$ with
respect to the bulk value.

We determined the temperature dependency of the Josephson critical
current by fitting the experimental $I-V$ curves with the
resistively shunted junction (RSJ) model. The results give
evidence of the superconductor-normal metal-superconductor long
weak link nature of the Josephson break junctions. Finally, by
applying to the junctions a microwave excitation, we observed
clear Shapiro steps and we were able to study the dependency of
their height on the normalized MW voltage.

The starting MgB$_2$ polycrystal bulk material, of cylindrical
shape, was obtained by reaction sintering of elemental B and Mg
for 3 h  at 950 $^\circ$C in a sealed stainless steel container,
lined with Nb foil. Details about the preparation technique will
be given elsewhere \cite{ref12}. The resulting highly dense
(density = 2.40 g/cm$^3$) MgB$_2$ cylinders had a metallic
appearance and a very high hardness. By using a fine, 0.15 mm
thick diamond circular saw we cut from the bulk material small
thin plates ($\sim 2 \times 1$ mm$^2$) with a thickness of 0.2-0.5
mm. The polished surfaces of these plates presented a shiny
metallic aspect and, when observed by a metallographic microscope,
they showed large (up to 50-60 $\mu$m) single-crystal-like grains,
with dark specular surface, embedded in a more amorphous
metallic-like background.

The superconducting properties of these samples were studied by AC
susceptibility (field equal to 0.1 gauss at 10 kHz) and
resistivity (four-probe AC technique with frequency equal to 133
Hz) measurements. The main graph of Fig.~1 shows the resistivity
of our MgB$_2$ samples, while in the upper and lower insets the
real part of the susceptibility and an enlarged view of the
resistive transition are presented, respectively. Both the
measurements give a critical temperature
$T_\mathrm{c}=$ 38.8 K with a transition width $\Delta%
T_\mathrm{c}=$ 0.5 K as determined from the derivative of the real
part of the susceptibility. The low value of the residual
resistivity ($\rho_\mathrm{0}\approx$ 4 $\mu\Omega$cm) and the
sharpness of the resistive and diamagnetic transitions prove the
high quality of the samples.

\begin{figure}[t]
\begin{center}
\includegraphics[keepaspectratio,width=0.9\columnwidth]{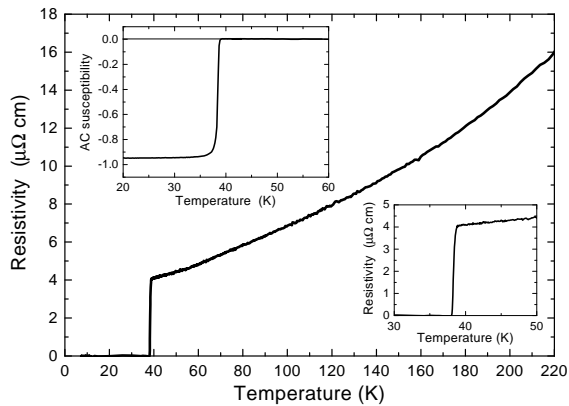}
\vspace{3mm}\caption{\small{Temperature dependency of the
resistivity of a MgB$_2$ sample; upper inset: temperature
dependency of the real part of the AC susceptibility of the same
sample; lower inset: enlarged view of the resistive
transition.}}\vspace{3mm}
\end{center}
\end{figure}

The thin MgB$_2$ plates were successively cut with the diamond saw
into small parallelepiped samples ($\sim$0.5$\times 0.5 \times 2$
mm$^{3}$) which were then fixed to an insulated elastic support by
means of epossidic glue. The current and voltage contacts on the
samples were obtained by means of Ag paste and showed a contact
resistance of the order of 0.2~-~1~$\Omega$. By breaking the
sample at 4.2~K and then carefully adjusting the pressure between
the two parts by means of a tip which bends the elastic sample
holder \cite{ref13}, we were able to obtain reproducible and
stable non-hysteretic Josephson junctions with normal resistances
between $\sim$ 0.1 and 11~$\Omega$. The $I-V$ curves were recorded
by injecting a current in the junction (Keithley 220 current
source) and measuring the voltage drop at its ends (HP 3457A
digital voltmeter).

An example of the low-temperature $I-V$ characteristics is shown
in Fig.~2 (main graph). In this case at $T =$ 5.28 K we have
$I_\mathrm{c} \approx 7$ mA and $R_\mathrm{N} \approx$ 0.15
$\Omega$ from which a product $I_\mathrm{c}R_\mathrm{N} \approx 1$
mV is obtained. This value is very small if compared to the value
$I_\mathrm{c}R_\mathrm{N}= \pi \Delta / 2 = 3.52 \pi k_\mathrm{B}
T_\mathrm{c} / 4 \approx 9.3$ mV predicted by the BCS theory for
$T_\mathrm{c}\approx 39$ K.

\begin{figure}[t]
\includegraphics[keepaspectratio,width=0.9\columnwidth]{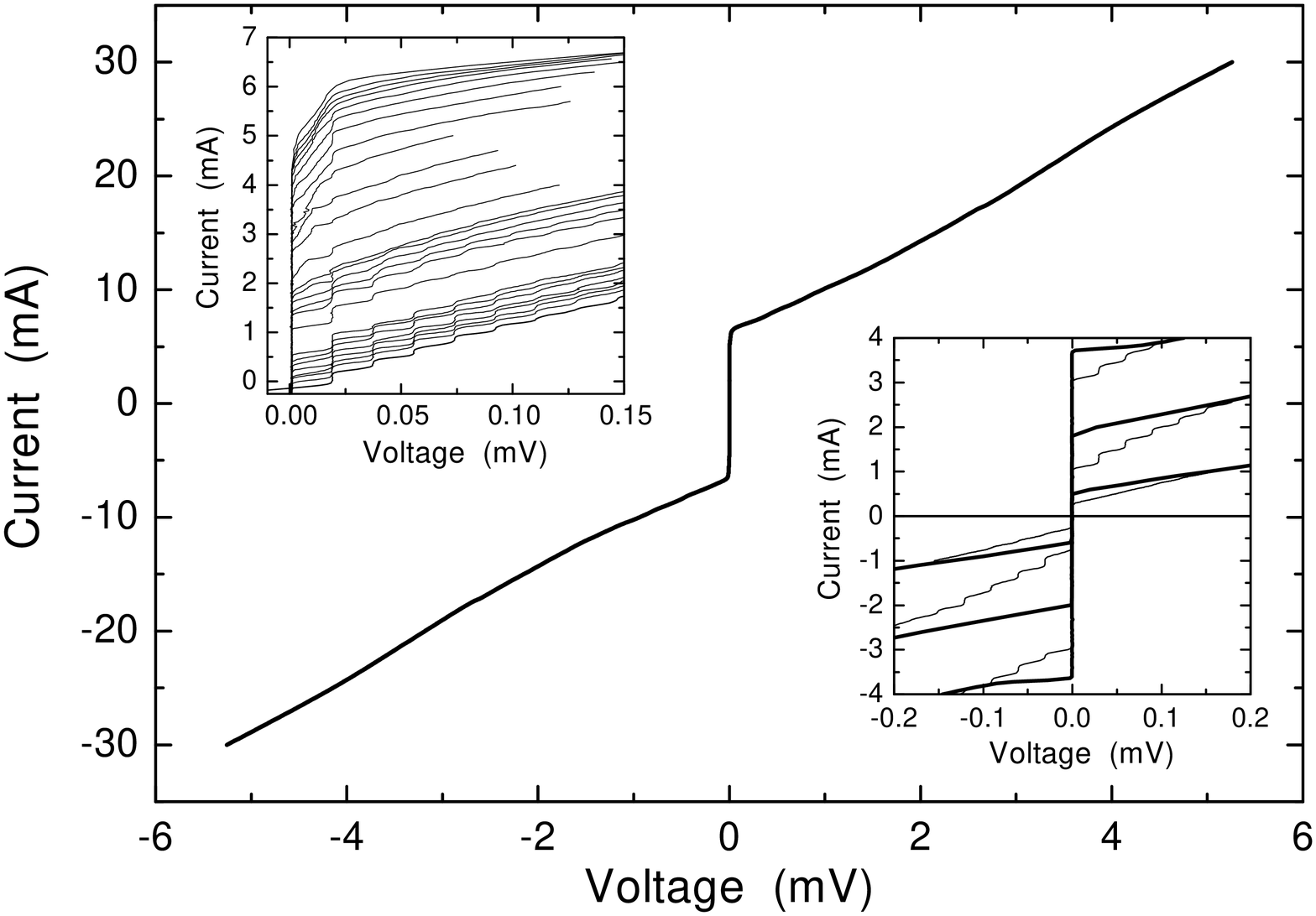}
\vspace{2mm} \caption{\small{Current vs. voltage characteristic of
a MgB$_2$ break junction at 5.28 K. In the upper inset, the $I-V$
curves of the same junction for increasing rf powers (from zero to
20 mW going from top to bottom) of an applied microwave radiation
at 8.843 GHz are shown. The lower inset shows the $I-V$ curves of
three other junctions with and without microwave excitation at
14.58 GHz (thin and thick solid lines, respectively).}}
\vspace{2mm}
\end{figure}

By applying to this junction a microwave (MW) excitation at $\nu =
8.843$ GHz, clear Shapiro steps spaced by the theoretical value
$\Delta V = h \nu / 2 \mathrm{e} \approx$ 18.3 $\mu$V were
observed, whose heights modulated at the change of the MW power.
These current steps were observed at voltages up to 1.1~mV, and
for some values of the MW power sub-multiples Shapiro steps at $V
= nh \nu /4 \mathrm{e}$ were also present. The curves reported in
the upper inset of Fig.~2 were obtained in the same junction of
the main graph, by varying the MW power from zero up to the
maximum power available from our source (20 mW). The presence of
the steps and their spacing in voltage are a direct proof of the
Josephson nature of the zero-bias current. By varying the contact
and changing the samples we obtained various $I-V$ curves with the
same features of that shown in the main graph of Fig.~2. In the
lower inset of the same figure, for example, the characteristics
at 4.2 K of three different junctions both with and without MW
excitation at 14.58 GHz are presented. The shape of all these
curves (when no RF signal is applied) is similar to that observed
in superconductor-normal metal-superconductor (S-N-S) junctions
\cite{ref14,ref15,ref16} and close to the prediction of the
standard RSJ model for a current-biased junction and in the
presence of thermal fluctuations. We will discuss this point in
the following.

All the results reported so far were obtained in junctions having
$R_\mathrm{N}\lesssim 1 \Omega$ and a high critical current
(\makebox{$I_\mathrm{c}\sim 0.6-8$ mA}). In these junctions the
$I_\mathrm{c}R_\mathrm{N}$ value remains small (0.3-1.7 mV) and
all the $I-V$ curves present the same features (included the AC
Josephson effect) as those shown in Fig.~2. We also obtained a
second kind of junctions, having $R_\mathrm{N}> 1 \Omega$ (up to
$\sim 11 \Omega$) and a smaller zero-bias current ($I_\mathrm{c}
\sim 0.3-2.5$ mA). In these junctions the product
$I_\mathrm{c}R_\mathrm{N}$ reaches the largest values
($3.2-3.8$~mV) and the $I-V$ characteristics are still
non-hysteretic, but present some differences with respect to those
of the first kind. In particular an evident knee is present at
$\mid V \mid \approx 4-5$ mV, that we interpret as due to the
quasiparticle tunneling features of the junction. Some traces of
this knee (always in the same position) are also present in most
of the $I-V$ curves of the first kind at low temperature (see
Fig.~2). From the local peaks of the junction's dynamic
conductance d$I$/d$V$ due to these knees, an approximate value of
the low temperature gap $\Delta \sim 1.7-2$ meV can be obtained.
Unfortunately, applying the MW excitation to the junctions with
higher $I_\mathrm{c}R_\mathrm{N}$ causes the total cancellation of
the zero-bias current, without producing any sign of Shapiro
steps. This result proves the non-Josephson nature of the
zero-bias current in the higher-resistance junctions. In the
present paper we only analyze the $I-V$ curves of the junctions
showing both the DC and AC Josephson effect.

\begin{figure}[t]
\vspace{1mm}
\begin{center}
\includegraphics[keepaspectratio,width=0.8\columnwidth]{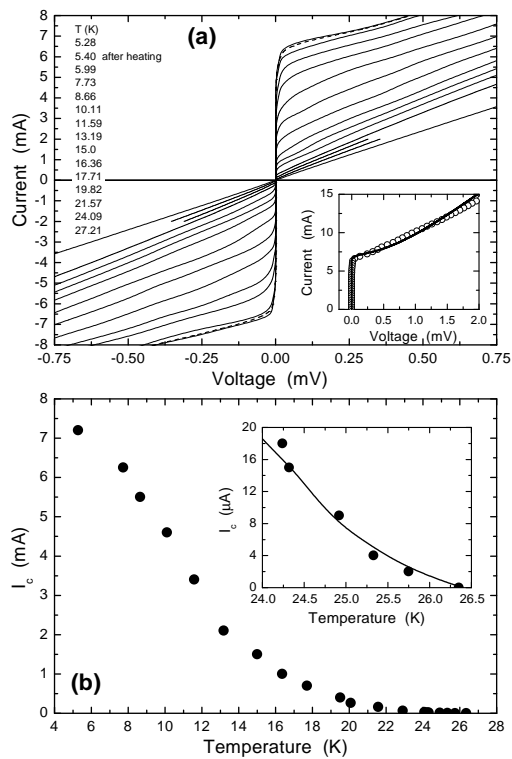}
\vspace{2mm} \caption{\small{(a) Temperature dependency of the
$I-V$ curves of the same contact shown at low $T$ in Fig.2. The
inset shows the RSJ fit (solid line) of the data at 5.28 K (open
circles). (b) Temperature dependency of the critical current and
low-temperature fit by the model for S-N-S weak link structures
(solid line in the inset).}}
\end{center} \vspace{2mm}
\end{figure}

The mechanical stability of the junctions during thermal cycling
was usually high. As a consequence, we were able to determine the
temperature dependency of the $I-V$ curves in most of the
contacts. Fig.~3(a) shows this dependency between $T=$ 5.28 K and
the \emph{critical temperature of the junction}
$T_\mathrm{c}^\mathrm{j} \approx 26-27$~K for the same contact
already shown at low $T$ in Fig.~2. In this figure, for clarity,
only some of the measured curves are presented. To give an idea of
the contact stability, we also report in Fig.~3(a) the $I-V$
characteristic measured at 5.4 K after the heating of the break
junction up to 40~K (dashed line). Very small differences are
present with respect to the curve measured before the thermal
cycle.

Even though, as we will see, the intrinsic nature of our junctions
is related to the presence of S-N-S weak links, as a first
approximation we can compare the $I-V$ curves with the prediction
of the RSJ model \cite{ref16} for current-biased junctions with a
very small capacitance and in the presence of thermal
fluctuations. The result of this comparison is shown in the inset
of Fig.~3(a) for the $I-V$ curve at $T = 5.28$\@K  (open circles
are experimental data while the solid line is the result of the
RSJ model). By applying this approach to all the data shown in
Fig.~3(a) we determined the $I_\mathrm{c}$ vs. $T$ dependency that
is presented in Fig.~3(b). The expanded view around
$T_\mathrm{c}^\mathrm{j} = 26.4$ K reported in the inset of the
same figure (solid circles) clearly shows that, at $T \approx
T_\mathrm{c}^\mathrm{j}$, $I_\mathrm{c} \propto
(T_\mathrm{c}^\mathrm{j}-T)^2$. This is a sign of the S-N-S nature
of the contact in our break junctions. A comparison with the
predictions of the De Gennes theory for the proximity effect in
metal barrier junctions \cite{ref16} gives good results at 22 K $<
T < T_\mathrm{c}^\mathrm{j}$ but completely fails at low
temperature.

A temperature behaviour of the critical current very similar to
our experimental data in the whole temperature range is instead
predicted by the theory of weak links with S-N-S structure, which
was derived by solving the Usadel equations for a one-dimensional
structure with electrodes in equilibrium and a zero critical
temperature in the weak link material \cite{ref17,ref18}. The
$I\ped{c}$ vs $T$ theoretical curve is similar to the experimental
one shown in Fig.~3(b) for values of the parameter $L /%
\xi_\mathrm{N}(T_\mathrm{c})$ between 6 and 10. Here $L$ is the
length of the normal weak link and $\xi_\mathrm{N}(T_\mathrm{c})$
is the decay length in the weak link material at the critical
temperature of the banks. The solid line in the inset of Fig.~3(b)
represents the theoretical curve predicted by this model at $T
\approx T_\mathrm{c}^\mathrm{j}$ which best agrees with our
experimental data. It was calculated by using
$T\ped{c}=T_\mathrm{c}^\mathrm{j} = 26.4$\@K and the corresponding
BCS gap $\Delta = 4$ meV, and adjusting the value of $L /
\xi_\mathrm{N}(T_\mathrm{c})$ to get the best fit (obtained for $L
/ \xi_\mathrm{N}(T_\mathrm{c})=9.9$). Nevertheless, other values
of the parameters, ranging between $\Delta = 1.7$ meV, $L /
\xi_\mathrm{N}(T_\mathrm{c}) = 7.85$ and $\Delta = 4.2$ meV, $L /
\xi_\mathrm{N}(T_\mathrm{c}) = 10$, give theoretical curves in
good agreement with the experimental data at $T > 0.9$\@
$T_\mathrm{c}^\mathrm{j}$.

The present theory for SNS weak link structures is valid only for
a superconductor in the dirty limit. This condition is not
fulfilled in the \emph{bulk} MgB$_2$ since $\ell = 600 $\AA and
$\xi \approx 50 $\AA~\cite{ref19}. Nevertheless we argue that the
conditions for the dirty limit can be fulfilled, at least at low
temperature, in the region of our contact: the low
$T_\mathrm{c}^\mathrm{j}$ with respect to the bulk $T_\mathrm{c}$
of MgB$_2$ indicates a local depression of the superconducting
properties in the bank material with a possible increase of
disorder and a reduction of $\ell$ in the weak link region. At
least two explanations are possible for this low
$T_\mathrm{c}^\mathrm{j}$ value: i) an intrinsic layer with
depressed superconducting properties at the surface of the MgB$_2$
grains; ii) an extrinsic effect of damaging maybe due to the great
mechanical stress produced in the region where the break has
occurred. On the other hand we have a clear evidence of a
reduction of $T_\mathrm{c}^{j}$ from $\approx$~33~K to
$\approx$~26~K after repeated thermal cycling of the same junction
between 4.2~K and room temperature, even though very good
Josephson characteristics are still present a few days after
breaking.

\begin{figure}[t]
\vspace{3mm}\hspace{5mm}
\includegraphics[keepaspectratio,width=0.8\columnwidth]{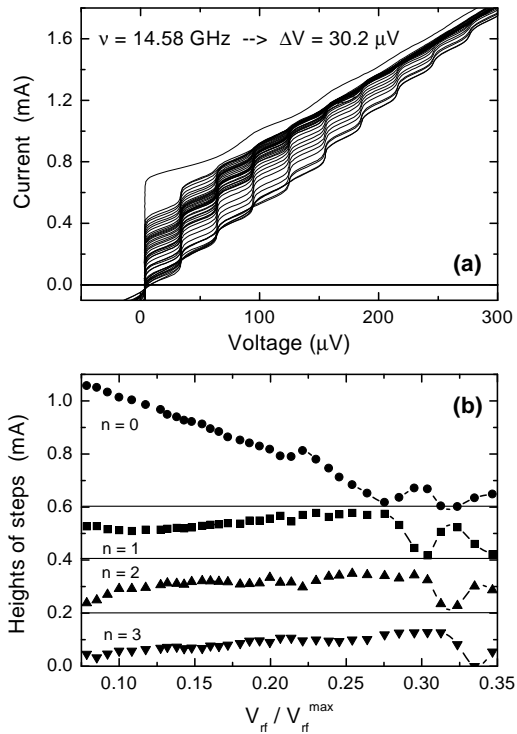}
\vspace{2mm} \caption{\small{(a) Shapiro steps of a break junction
in the presence of external microwave radiation at $\nu = 14.58$
GHz at different power levels. The upper curve is at zero MW
power; (b) Magnitude of the supercurrent and the first three
current steps as function of the normalized rf
voltage.}}\vspace{8mm}
\end{figure}

Finally, in Fig.~4(a) and (b) we present in greater detail the AC
Josephson effect results. Fig.~4(a) shows a set of $I-V$ curves
measured in a junction with low resistance after irradiation with
microwaves at $\nu$ = 14.58 GHz and various powers. We observed up
to 9~-~10 vertical Shapiro steps spaced by the ideal value $\Delta
V$ = 30.2 $\mu$V.

The magnitude of the supercurrent ($n=0$) and the heights of the
first three steps ($n = 1, 2, 3$) as functions of the normalized
RF voltage are shown in Fig.~4(b). The frequency of the applied
microwave gives a normalized frequency $\Omega = h \nu / 2
\mathrm{e}R_\mathrm{N}I_\mathrm{c} = 0.19$. The experimental
curves shown in Fig.~4(b) are in good qualitative agreement with
the results of the standard RSJ current biased model of the
irradiated junction \cite{ref16}.

In conclusion, the first experimental evidence of the Josephson
effect in MgB$_2$ is given by using the break-junction technique.
This result is a direct prove for the existence of pairs with
charge $2e$ in this new metallic superconductor. The product
$I_\mathrm{c}R_\mathrm{N}$ of our junctions is small if compared
to the BCS value but critical currents up to 8~mA were observed.
The temperature dependency of the critical current gives evidence
for a S-N-S long weak link nature of the junctions and for a
depressed $T_\mathrm{c}$ in the region of the contact.

Many thanks are due to A. Barone, A. Bianconi and A. Perali for
useful discussions. One of the authors (V.A.S.) acknowledges the
partial support by the Russian Foundation for Basic Research
(grant No 99-02-17877) and by the Russian Ministry of Science and
Technical Policy within the program ``Actual Problems of Condensed
Matter Physics'' (grant No  96001). \vspace{3mm}

%
\end{document}